\documentclass[%
 reprint,
superscriptaddress,
 amsmath,amssymb,
 aps,
]{revtex4-2}

\usepackage{graphicx}
\usepackage[dvipsnames]{xcolor}
\usepackage{dcolumn}
\usepackage{bm}
\usepackage[]{hyperref}
\usepackage{url}
\usepackage{cancel}
\usepackage{soul}
\usepackage{ulem}
\usepackage{xcolor}
\usepackage{tikz}

\begin{document}


\title{A centimeter-sized gas pressure sensor for high-vacuum measurements \\ 
at cryogenic temperatures}

\author{Christoph Reinhardt}%
\affiliation{Deutsches Elektronen-Synchrotron DESY, Notkestr. 85, 22607 Hamburg, Germany}

\author{Lea Lara Stankewitz}
\email{Current address: Institut für Experimentalphysik, Universität Hamburg, 22761 Hamburg, Germany}
\affiliation{Deutsches Elektronen-Synchrotron DESY, Notkestr. 85, 22607 Hamburg, Germany}
\affiliation{Institut für Quantenphysik (IQP) \& Zentrum für Optische
Quantentechnologien (ZOQ), Universität Hamburg, 22761 Hamburg, Germany}

\author{Daniel Hartwig}%
\affiliation{Deutsches Elektronen-Synchrotron DESY, Notkestr. 85, 22607 Hamburg, Germany}

\author{Sandy Croatto}
\affiliation{Deutsches Elektronen-Synchrotron DESY, Notkestr. 85, 22607 Hamburg, Germany}

\author{Hossein Masalehdan}%
\email{Current address: Fraunhofer-Institut für Siliziumtechnologie ISIT, Fraunhoferstraße 1, 25524 Itzehoe, Germany}
\affiliation{Institut für Quantenphysik (IQP) \& Zentrum für Optische
Quantentechnologien (ZOQ), Universität Hamburg, 22761 Hamburg, Germany}

\author{Nils Sültmann}%
\affiliation{Institut für Quantenphysik (IQP) \& Zentrum für Optische
Quantentechnologien (ZOQ), Universität Hamburg, 22761 Hamburg, Germany}

\author{Axel Lindner}%
\affiliation{Deutsches Elektronen-Synchrotron DESY, Notkestr. 85, 22607 Hamburg, Germany}

\author{Roman Schnabel}%
\affiliation{Institut für Quantenphysik (IQP) \& Zentrum für Optische
Quantentechnologien (ZOQ), Universität Hamburg, 22761 Hamburg, Germany}


\begin{abstract}
Gas pressure sensors based on nanomechanical membranes have recently demonstrated an ultra-wide ten-decade measurement range, a gas-type-independent response, and a self-calibrating operation with uncertainties of approximately $1\,\%$. 
The readout relied on tabletop free-space laser interferometers. 
Here we present a centimeter-sized, portable implementation in which a square Si$_3$N$_4$ membrane is read out via a fiber-based laser interferometer. 
We perform pressure measurements between $5\times10^{-5}$ and $10^{-1}$~mbar in a confined $0.7$~L volume cooled to $78$~K. 
Because no suitable commercial pressure sensor exists for direct cryogenic comparison, we benchmark our device against room-temperature commercial gauges connected to the cold volume through a pipe of limited conductance. 
The measured relationship between the two sensors is compared with models accounting for temperature- and pumping-induced pressure gradients within the measurement chamber. 
These models agree with the measurements to within $<10\,\%$ for helium and $<13\,\%$ for nitrogen.
The achieved readout sensitivity of $S_x = 8\times10^{-14}\,\mathrm{m}/\sqrt{\mathrm{Hz}}$ theoretically enables resolving the thermal displacement noise spectrum of a trampoline membrane at atmospheric pressure, with a peak response of $48\,S_x$ $\left(25\,S_x\right)$ at $295\,\mathrm{K}$ $\left(78\,\mathrm{K}\right)$. 
Our results suggest that the previously achieved pressure measurement range of ten decades with trampoline membranes is compatible with fiber-based optical readout. 
This paves the way for widely applicable pressure sensors in the centimeter size range in 
cryogenic environments.
\end{abstract}

\maketitle

\section{Introduction}
Gas-induced damping of a mechanical oscillator provides a well-established method for measuring the pressure of a gas or fluid \cite{christian1966theory}.
Recent work has extended this principle to chip-scale mechanical oscillators, demonstrating ultra-compact form factors, scalable fabrication, high sensitivity, and robust operation in harsh environments \cite{chen2022nano,bargatin2012large,dolleman2016graphene,patel2018dual,huisman2014quartz,li2007ultra}. Within this class, gas pressure sensors that integrate nanomechanical resonators with laser-interferometric readout have emerged as particularly promising \cite{blom1992dependence,bianco2006silicon,verbridge2008size,dolleman2016graphene,naesby2017effects}.
The development of ultra-low-loss thin-film resonators made of silicon nitride or silicon carbide has recently enabled significant advances in this field:
These include a device with a measurement range spanning 10 orders of magnitude \cite{reinhardt2024self}, a sensor capable of direct pressure measurements independent of gas type \cite{salimi2024squeeze}, and demonstrations achieving approximately 1~\% measurement uncertainty in alignment with primary pressure standards \cite{green2025accurate}.

Although the sensing element was realized on-chip, previous optical readout of nanomechanical pressure sensors relied on free-space laser interferometers, resulting in a table-top-scale setup \cite{reinhardt2024self,salimi2024squeeze,green2025accurate}.
To enable portability and broad applicability of these sensors, a compact readout is required.
Several optical and electrical approaches are potential candidates to this end, including compact fiber-based interferometers \cite{rugar1989improved}, on-chip optical resonators \cite{li2021cavity,guo2022integrated}, capacitive coupling to LC circuits \cite{truitt2007efficient,bagci2014optical,puglia2025room}, magnetomotive readout \cite{cleland1998nanometre,chien2020nanoelectromechanical}, and piezoelectric transduction \cite{o2010quantum,ciers2024nanomechanical}.
Compared with electrical approaches, optical readout avoids depositing conductive layers on the resonator, eliminating non-uniform mass loading, ohmic losses, and other factors that can degrade sensitivity and accuracy.
 
Realizing a cm-sized, high-vacuum-compatible \mbox{($<10^{-3}$~mbar)} nanomechanical pressure sensor with fiber-based optical readout not only enhances its use in typical vacuum systems but also enables local pressure measurements in constrained volumes and at cryogenic temperatures. 
Particularly at cryogenic temperatures, commercial sensors, specified for temperatures down to 78~K, are generally limited to pressure ranges well above $10^{-3}$~mbar (e.g., Kulite CTL-190(M) \cite{Kulite_CTL190}).
Pressure monitoring at low temperatures is essential for quantum devices, particle physics experiments, gravitational-wave detection, and energy storage \cite{radebaugh2016cryogenic, arpaia2018state,bahre2013any,franke2024measurement}.
Demonstrated solutions include capacitive, piezoresistive, quartz, and ionization-based gauges \cite{lago2014rugged, nara1993piezo, swanson1998accurate, huisman2014quartz, lotz2014development}, and, more recently, implementations based on micro-electromechanical systems (MEMS) \cite{nguyen2018highly, verma2020sensitivity, zhao2024wireless}.
Making the ultra-wide measurement range, calibration-free accuracy, and gas-type-independent response of recent nanomechanical pressure sensors \cite{reinhardt2024self,salimi2024squeeze,green2025accurate} accessible under cryogenic conditions would constitute a significant advance with regard to demonstrated solutions.

Here, we demonstrate a portable, centimeter-sized assembly that integrates a nanomechanical resonator with a fiber-based optical interferometer for displacement readout.
The resonator consists of a low-mechanical-loss, square membrane fabricated from silicon nitride (Si$_3$N$_4$), which serves as the sensing element.
We use this device to perform local pressure measurements between $5\times10^{-5}$ and $10^{-1}$~mbar in a confined $0.7$~L volume cooled to $78$~K.
Because no commercial cryogenic pressure sensor is available for direct comparison, we benchmark our device against room-temperature gauges connected to the cold volume through a pipe of limited conductance, and we apply both analytical and finite-element models to account for pressure gradients arising from temperature gradients (i.e., thermal transpiration) and active pumping.
The relative deviation between measured and modeled pressures remains within $10\,\%$ for helium and within $13\,\%$ for nitrogen.
The realized displacement readout sensitivity of $S_x = 8\times10^{-14}\,\mathrm{m}/\sqrt{\mathrm{Hz}}$ is sufficient to resolve the expected thermal displacement noise of a trampoline resonator at atmospheric pressure \cite{reinhardt2024self}, corresponding to peak responses of $48\,S_x$ at 295~K and $25\,S_x$ at 78~K.  
This indicates that the demonstrated fiber-interferometric readout is compatible with the ten-decade pressure measurement range previously achieved using trampoline membranes \cite{reinhardt2024self}, thereby providing a viable route toward broadly applicable, centimeter-sized nanomechanical pressure sensors.

\section{Sensor assembly and readout sensitivity}
\label{sec:sensitivity}
\begin{figure}
\includegraphics[width=\linewidth]{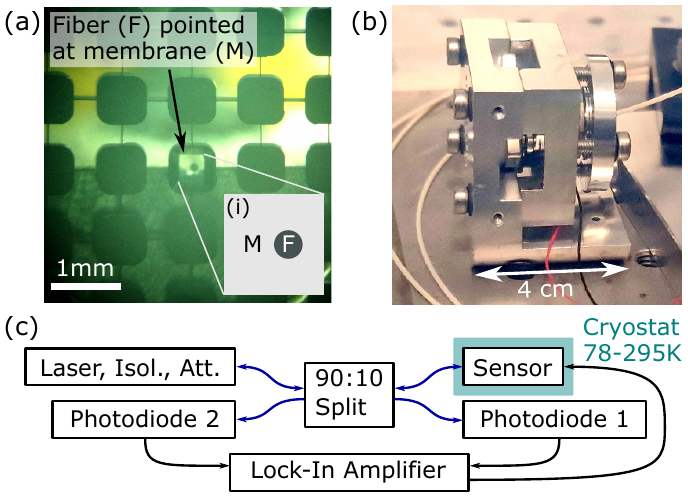}
\caption{Nanomechanical gas pressure sensor with fiber-optical readout.
(a) Microscopic image of the square Si$_3$N$_4$ membrane (M) with 350~\textmu{}m side length, which is suspended from a periodically-patterned silicon chip for acoustic isolation.
A cleaved optical fiber of circular cross section (F) is placed at a distance of 160~\textmu{}m behind the membrane (M), as schematically shown in inset (i) thereby forming a Fabry-Perot interferometer.
(b) Photograph of the pressure sensor assembly, with dimensions of approximately 4~cm in each direction.
The assembly includes two piezoelectric actuators: one for setting the fiber–membrane separation, contacted by red and black cables, and one for membrane actuation (located on the back side). The optical fiber with its white buffer coating is also visible.
(c) Schematic of the fiber-coupled interferometric readout. Optical (electrical) signals are indicated by blue (black) arrows. Laser light at 1550~nm is directed to the sensor inside a cryogenically cooled vacuum chamber (78–295~K), and the back-reflected signal is detected with a photodiode and lock-in amplifier to read out the membrane motion. The lock-in amplifier output is applied to a piezoelectric actuator to excite membrane oscillations.
}
\label{fig:fiber_sensor}
\end{figure}

Figure~\ref{fig:fiber_sensor}(a) shows the sensing element corresponding to an acoustically-shielded high-stress Si$_3$N$_4$ square membrane \cite{yu2014phononic,tsaturyan2014demonstration}.
It has a side length of 350~µm, a tensile stress of 0.9~GPa, and a thickness of $d=40$~nm.
An optical SMF28 fiber (cladding diameter 125~µm) is aligned perpendicular to the membrane close to its center, thereby forming a fiber interferometer \cite{rugar1989improved}.
Using a microscope, the fiber-membrane separation was determined to be approximately 160~µm.

Figure~\ref{fig:fiber_sensor}(b) shows a side view of the sensor assembly.
Here, the membrane chip is clamped all around its perimeter, to a metal piece (left), using a metal clamp with an interjacent Polytetrafluoroethylene shim.
The optical fiber (having a white plastic buffer) is attached to a second metal piece (right) using cryo-compatible epoxy resin (Loctite Stycast 2850FT), which in turn is connected to the membrane holder via a third piece (center).
A ring-shaped piezoelectric actuator (Physik Instrumente PD150.3x) between the fiber holder and the central metal piece enables scanning the fiber-membrane separation over a 1.5~µm distance, which enables setting the interferometer to the mid-fringe position.
Another piezo actuator (Physik Instrumente PD050.3x) is attached to the back side of the membrane holder, enabling gas-pressure-dependent measurements of the mechanical quality factor ($Q$) by recording the ringdown of the oscillation amplitude after stopping resonant excitation \cite{reinhardt2024self} (see Sec.~\ref{sec:P_measurement} for details).  
All metal parts are fabricated from aluminum.

Figure~\ref{fig:fiber_sensor}(c) shows a schematic of the experimental setup for reading out the fiber-coupled membrane pressure sensor.
The sensor is mounted inside a vacuum chamber with adjustable pressures $P$ between $10^{-6}$ and 1~mbar and temperatures $T$ between 78 and 295~K (Fig.~\ref{fig:cryostat}(a)).
A 1550~nm laser (Thorlabs ULN15TK) passes through an isolator, is attenuated by 20~dB, and is split in a 90:10 ratio.
Ninety percent of the light is directed to photodiode 1 (Thorlabs PDA05CF2) for monitoring intensity stability, using the auxiliary input of a lock-in amplifier (LIA; Zurich Instruments HF2).
The remaining 10~\%, corresponding to an input power of $P_\mathrm{in}=1.3\,\mathrm{mW}$, is sent to the membrane, which is excited to oscillate by applying the LIA's output voltage to the corresponding piezo.
The back-reflected signal of power $P_\mathrm{r}$ is amplitude-modulated by the oscillating membrane.
It is split such that 90~\% is detected on photodiode 2 (Thorlabs PDA05CF2), and subsequently processed via the LIA input, which provides amplitude and phase information.
The membrane's resonance frequency is obtained by sweeping the excitation frequency and identifying the maximum response.
The remaining 10~\% of the back-reflected light reaches the isolator, preventing propagation towards the laser.

The displacement sensitivity is determined by how efficiently membrane-reflected light is coupled back into the fiber.
Only the fraction of the reflected field matching the fiber eigenmode is recaptured.
The main sources of loss are (i) angular misalignment between the fiber tip and membrane, and (ii) free-space mode divergence, which grows with increasing fiber–membrane separation.
To quantify these effects, we performed plane-wave spectrum simulations of the optical field propagation~\cite{li_theoretical_2006, reinhardt_ultralow-noise_2017}, decomposing the optical modes into plane waves and propagating them through the geometry.
Multiple reflections between the fiber and membrane facets [Fig.~\ref{fig:optics_sim}(a)] were included.
Further details are given in Appendix~\ref{appsec:plane_wave_spectrum}.

\begin{figure}
\includegraphics[width=\linewidth]{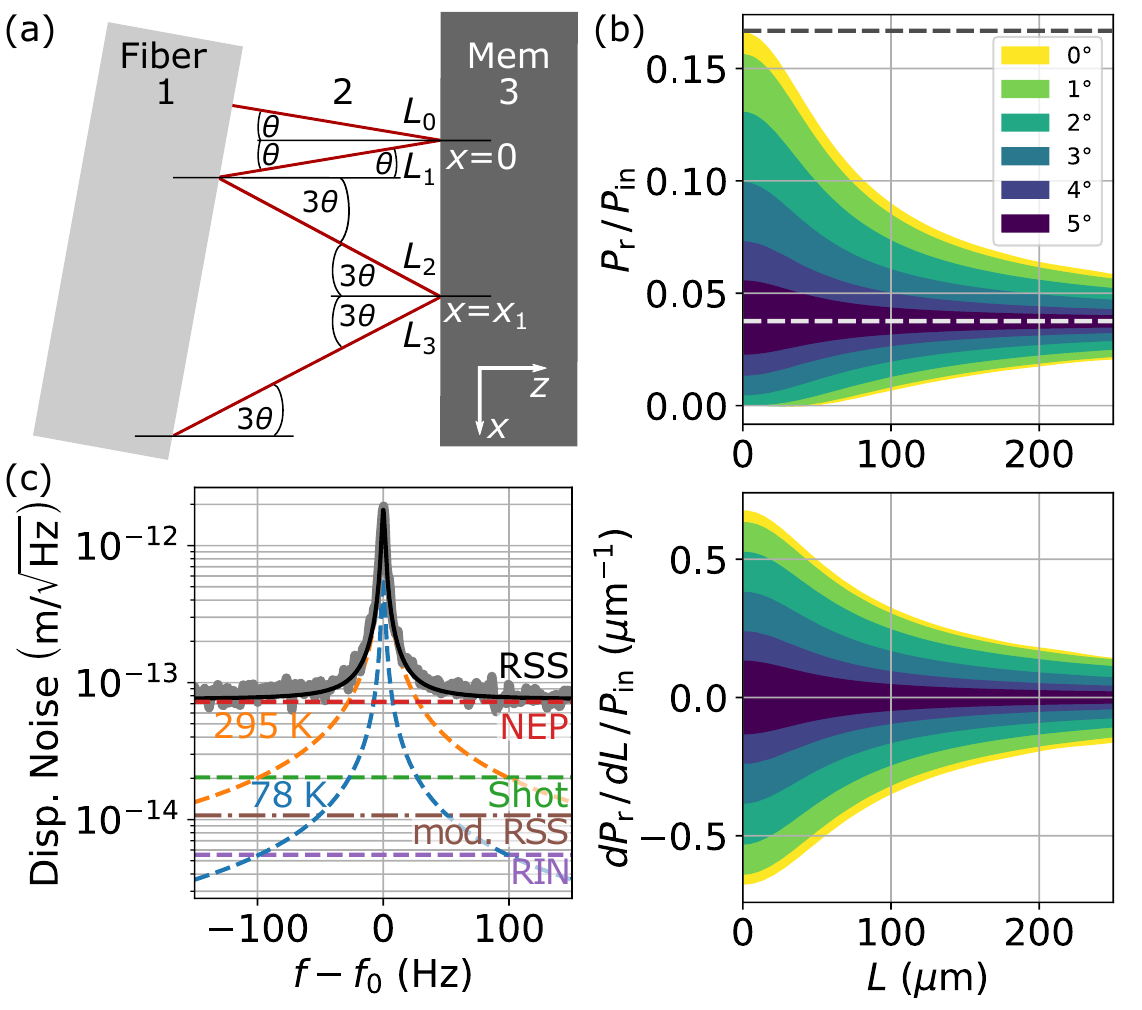}
\caption{Readout sensitivity of the fiber interferometer.
(a) Angular misalignment $\theta$ between the fiber tip and the membrane (Mem) leads to an increasing lateral offset and mode mismatch with each reflection between the fiber and the reflected light.
(b, upper) Reflected light power fraction as a function of fiber–membrane distance~$L$ for different misalignment angles.
Colored surfaces indicate the extent of the interference fringe; the dashed white (gray) line marks the fiber tip reflectivity (plane-wave Fabry–Pérot limit, Eq.~\ref{eq:plane_wave_limit}).
(b, lower) Gradient of the reflected power fraction with respect to displacement~$L$, which defines the interferometric responsivity.
(c) Measured displacement noise of the membrane at its intrinsic $Q$ factor [see Fig.~\ref{fig:cryostat}(b)] at 295~K (gray). 
Modeled thermal displacement noise of the membrane at 295~K and 78~K is shown in orange and blue, respectively. 
Also shown are contributions from the photodetector noise-equivalent power (NEP, red), shot noise (green), and laser relative intensity noise (RIN, purple). 
The black curve (RSS) denotes the root-sum-square of all individual noise contributions for the demonstrated sensor ($L=160$~µm, $\theta=2.5^\circ$), while the brown curve (RSS mod.) indicates the predicted readout noise level for an optimized sensor (see text).}
\label{fig:optics_sim}
\end{figure}

Figure~\ref{fig:optics_sim}(b) shows the reflected power $P_\mathrm{r}$ (upper) and its derivative (lower), both normalized to the input power $P_\mathrm{in}$, as a function of the fiber–membrane spacing $L$.
Colored areas represent the range between minima and maxima of the approximately sinusoidal interference fringe, for angular misalignments $\theta$ of $0\,–\,5^\circ$.
For misalignments above $1^\circ$ and separations beyond 20~µm, both the interference contrast and the interferometric responsivity $\mathcal{R}=\mathcal{T}\,\mathrm{d}P_\mathrm{r}/\mathrm{d}L$, where $\mathcal{T}=0.9$ is the transmittance towards Photodiode 2, decrease significantly, with respect to the plane-wave Fabry–Pérot limit (dashed gray line, Eq.~\ref{eq:plane_wave_limit}), which degrades the readout sensitivity.
The dashed white line marks the reflectance of the fiber tip, which the signal approaches for increasing $L$ and $\theta$.
These results were verified by finite-difference time-domain simulations~\cite{hartwig2025smsi}.

For the studied sensor, with a spacing of $L=160$~µm, the angular misalignment is consistent with $\theta\approx2.5^\circ$, corresponding to a reflectance gradient of about $14\,\%/$\textmu{}m.
From this we derive the equivalent displacement noise spectral densities of the relevant noise sources, which are shown together with the measured displacement noise in Fig.~\ref{fig:optics_sim}(c).
The total noise floor, calculated as the root sum square (RSS) of the individual contributors, is compared to the thermal displacement spectra at 295~K and 78~K, given by~\cite{saulson1990thermal}
\begin{equation}
    S^{x}_\text{th}(f) = \frac{k_B T f\mathrm{m}^2}{2 \pi^3 m Q f \left[\left(f_\mathrm{m}^2 - f^2\right)^2 + f^4 Q^{-2} \right]},
\label{eq:thermal_noise}
\end{equation}
where $k_B$ is the Boltzmann constant, $T$ the membrane temperature, $m$ its effective mass, and $f_\mathrm{m}$ and $Q$ are the resonance frequency and quality factor of the mode, respectively.
The nearly equal peak heights at both temperatures arise from the balance between decreasing thermal energy and the increasing $Q$ factor [see Fig.~\ref{fig:cryostat}(b)].

The dominant technical noise is the amplified photodiode's noise-equivalent power (NEP) $\sqrt{S^{P}_\text{NEP}(f)}=11.6\,\,\mathrm{pW/\sqrt{Hz}}$, corresponding to an equivalent displacement noise
\begin{equation}
    S^{x}_\text{NEP}(f) = S^{P}_\text{NEP}(f)/\mathcal{R}^{2}.
\label{eq:nep}
\end{equation}

By contrast, the fundamental sensing limit for coherent light is the photonic shot noise (SN),
\begin{equation}
    S^{x}_\text{SN} = 2 h c \eta \bar{P}/\lambda\mathcal{R}^{2},
\label{eq:shot_noise}
\end{equation}
where $h$ is Planck’s constant, $c$ the speed of light, $\eta=0.82$ the photodetector quantum efficiency, and $\bar{P}\approx 6$~\textmu{}W the mean optical power incident on the detector.

Relative intensity noise (RIN) of the laser, 
\begin{equation}
    S^{x}_\text{RIN}(f) = S_\text{RIN}(f) \bar{P}^2/\mathcal{R}^{2},
\end{equation}
with $S_\text{RIN}(f=f_\text{m}) = -150\:\text{dBc}/\text{Hz}$ for our laser, is
negligible, as shown in Fig.~\ref{fig:optics_sim}(c).

The overall displacement sensitivity of the present setup corresponds to the RSS of the individual noise contributions (black line).
Although its current level, $S_x = 8\times10^{-14}\,\mathrm{m}/\sqrt{\mathrm{Hz}}$, already allows resolving the thermally driven membrane motion (blue and orange curves), significant further improvements should be readily achievable.
Reducing the fiber--membrane spacing to approximately 20~\textmu{}m and maintaining a misalignment below $1^\circ$, in combination with using a lower-noise detector (e.g., Thorlabs PDA10CS2, with $\sqrt{S^{P}_{\mathrm{NEP}}(f)} = 3.7\,\mathrm{pW}/\sqrt{\mathrm{Hz}}$ \cite{ThorlabsPDA10CS2}), could enable a shot-noise-dominated displacement sensitivity of approximately $10^{-14}\,\mathrm{m}/\sqrt{\mathrm{Hz}}$.
Reducing the displacement noise directly benefits pressure sensing, in readout-noise-limited regimes, as it allows ringdown measurements with the same signal-to-noise ratio at lower piezo-excitation amplitudes, thereby improving power efficiency.
Moreover, at higher pressures ($P \gtrsim 10^{-1}$~mbar), the pressure $P$ is typically determined from the thermally driven noise spectrum of the membrane \cite{naesby2017effects,reinhardt2024self}.
In this regime, lowering the displacement noise floor extends the accessible pressure range toward higher values.
Interestingly, the realized displacement sensitivity $S_x$ is compatible with the 10-decade pressure-measurement range previously demonstrated for a trampoline membrane~\cite{reinhardt2024self}, as it enables resolving the trampoline’s expected thermal-displacement noise of $48\,S_x$ ($25\,S_x$) at atmospheric pressure for 295~K (78~K) \cite{reinhardt2024self}.

\section{Cryogenic Pressure Measurements}
\label{sec:P_measurement}

\begin{figure}
\includegraphics[width=\linewidth]{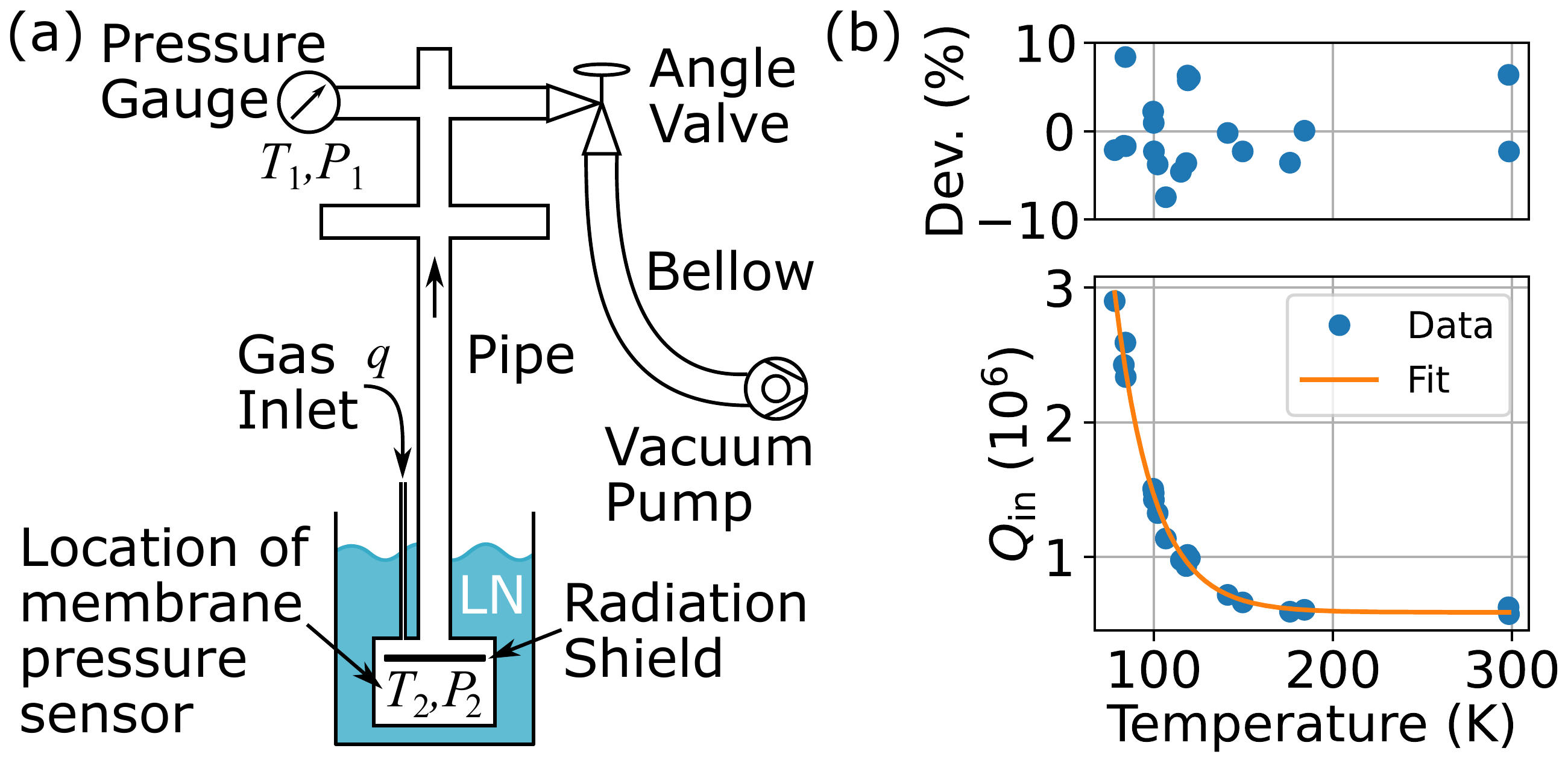}
\caption{Cryogenic vacuum setup for sensor characterization. (a) 
The vacuum chamber is mounted inside a liquid-nitrogen-filled (LN) dewar (78~K) and includes a thermal radiation shield, a gas inlet for He and N$_2$, and the membrane pressure sensor at the cold bottom. 
A thin pipe of diameter $D=35$~mm connects the cold volume to the room-temperature top, where a commercial pressure gauge provides reference readings. 
The system is evacuated through a bellow and angle valve. 
(b) Measurement (blue) of the membrane's intrinsic quality factor $Q_\text{in}$ versus temperature together with a fit function (orange; see text). 
The relative deviation between data and fit is shown above.}
\label{fig:cryostat}
\end{figure}

Figure~\ref{fig:cryostat}(a) shows a schematic of the home-built cryogenic measurement chamber, which is made out of stainless steel. 
The lower volume, containing our sensor of Fig.~\ref{fig:fiber_sensor}, is cooled down to approximately 78~K by immersion in a liquid nitrogen bath. 
A 500~mm long pipe of diameter $D=35$~mm thermally isolates lower and upper chamber volumes, such that the temperature gradient during cryogenic operation does not extend beyond the length of the pipe. 
A baffle mounted beneath the lower end of the pipe and above our sensor shields the sample volume against thermal radiation from the top. 
To adjust the pressure inside the chamber, gas is injected into the lower volume and pumped on the top via an angle valve and bellow.
Due to the limited gas conductance of the long and thin pipe (see Appendix~\ref{app:comsol}), pressure gradients between lower and upper parts of the vacuum chamber emerge, when pumping. 
If both volumes are at different temperatures, pressure gradients are additionally affected by thermal transpiration effects in the free molecular flow regime. 
Thermal transpiration describes the gas flow from the warm into the cold volume as a consequence of the $T$ dependency of the gas particles' mean free path $\ell = k_\mathrm{B} T/\left(\sqrt{2}\,\pi \delta^{2} P\right)$, with the effective diameter of a gas particle $\delta$ \cite{kennard1938kinetic}.

Figure~~\ref{fig:cryostat}(b) shows the intrinsic quality factor $Q_\mathrm{in}$ of the membrane’s fundamental mode ($f_\mathrm{m} = 1.1~\mathrm{MHz}$) as a function of temperature~$T$, measured at a base pressure of $P_\text{base}\sim 10^{-6}$~mbar. 
Each data point (blue circles) corresponds to $Q\left(P_2=P_\text{base}\right) = \pi f_\mathrm{m} \tau$, where the ringdown time~$\tau$ is obtained from an exponential fit to the amplitude decay following the interruption of a resonant piezo drive \cite{reinhardt2024self}. 
Upon cooling from 295~K to 78~K, the $Q$ factor increases by about a factor of five, from $0.6\times10^{6}$ to $2.9\times10^{6}$. 
The data agree within $\pm\,10\,\%$ with the fit model (orange line) describing dissipation due to thermally activated defects, $Q_\mathrm{in}^\mathrm{fit}\left(T_2\right) = \left[{1 + (2\pi f_\mathrm{m} \tau_\mathrm{rel})^2}\right]/\left[\sigma 2\pi f_\mathrm{m} \tau_\mathrm{rel}\right]$, where $\sigma$ characterizes the defect density and $\tau_\mathrm{rel} = \tau_{\mathrm{rel},0} \exp(E_A/k_\mathrm{B} T_2)$ follows the Arrhenius relation~\cite{IMBODEN201489}. 
The fitted parameters, $\sigma = 3.5\times10^{-6}$, $\tau_{\mathrm{rel},0} = 4.4\times10^{-8}$~s, and $E_A = 0.24$~eV, are consistent with previously reported values~\cite{IMBODEN201489,metcalf2018improving}.

We use our sensor to measure the pressure $P_2$ inside the lower volume of the cryogenic vacuum chamber shown in Fig.~\ref{fig:cryostat}(a), while observing the pressure $P_1$ in the upper volume with commercial pressure gauges (Leybold IONIVAC ITR 90, Pfeiffer CMR 365).
$P_2$ is determined according to \cite{reinhardt2024self} 
\begin{equation}
    P_{2} = f_\mathrm{m} \rho d \sqrt{\frac{\pi^3 k_b T_2}{8m_\mathrm{gas}}}\left[\frac{1}{Q\left(P_2\right)}+\frac{1}{Q_\mathrm{in}^\mathrm{fit}\left(T_2\right)}\right],
\end{equation}
with the mass $m_\mathrm{gas}$ of the gas particles, the membrane's mass density $\rho$, and its intrinsic quality factor $Q_\mathrm{in}=Q_\mathrm{in}^\mathrm{fit}$.

As a first step, we investigate the pressure gradient caused by thermal transpiration when cooling the lower part of the vacuum chamber to 78~K, while the upper part stays at 295~K. 
Prior to cooling, the chamber is evacuated and the inlet and outlet valves are closed. 
After cooling down, controlled amounts of gas are injected and the chamber is closed again. 
Once a stable pressure is established, we take five pressure measurements with our sensor. 
The resulting $P_2$ value corresponds to the arithmetic mean of these individual measurements, and the related standard error (i.e., the standard deviation of the sample mean) represents the random error in the mean, $\Delta P_{2,\mathrm{r}}$.

In the free molecular flow regime, defined by a Knudsen number $\mathrm{Kn}=\ell/D>10$, the relation $P_2/P_1=\sqrt{T_2/T_1}=1.94$ applies \cite{kennard1938kinetic}, which is confirmed by the COMSOL simulation shown in Fig.~\ref{fig:78K_pump_off}(a). 
These simulations make use of the heat transfer and molecular flow modules (see Appendix~\ref{app:comsol} for details). 
For viscous flow, the pressure equilibrates within the chamber, such that $P_2=P_1$.
The transition region between these two regimes is described by the semi-empirical Takaishi-Sensui (TS) model, \cite{tanuma2000ion,takaishi1963thermal} 
\begin{equation}
    P_{2,\mathrm{TS}} = \left[1+\frac{\sqrt{T_2/T_1}-1}{A(X/X_0)^2+BX/X_0+C\sqrt{X/X_0}+1}\right]P_1,
    \label{eq:TS}
\end{equation}
with $X_0=1\times10^{-5}$\,m\,mbar\,/\,K, $X=2DP_1/\left(T_1+T_2\right)$, and gas dependent parameters $A$, $B$, and $C$, given in Table~\ref{tab:ABC} for helium and nitrogen, respectively.   

\begin{table}[ht]
\centering
\begin{tabular}{c | c c c}
 & $A$ & $B$ & $C$ \\
\hline
Helium ($4.3$–$318$~K) \cite{tanuma2000ion} & 6.11 & 4.26 & 0.52 \\
\hline
Nitrogen ($77$–$195$~K) \cite{takaishi1963thermal} & 67.43 & 7.50 & 0.87 \\
\end{tabular}
\caption{Parameters $A$, $B$, and $C$, along with the specified temperature range, for Helium and Nitrogen, as used in the semi-empirical Takaishi-Sensui Model (Eq.~\ref{eq:TS}) describing thermal transpiration in the transitional flow regime.}
\label{tab:ABC}
\end{table}

The lower panels of Fig.~\ref{fig:78K_pump_off}(b) show the measured relations between $P_1$ and $P_2$ (green dots) together with the Takaishi–Sensui (TS) model predictions (green curves) for helium (left) and nitrogen (right).
In the upper panels, the corresponding relative deviations between measurements and model predictions, $\Delta P_{2,\mathrm{m}}/P_2 = (P_2 - P_{2,\mathrm{TS}})/P_2$, are indicated by $+$ symbols.
Relative random deviations, $\Delta P_{2,\mathrm{r}}/P_2$, are shown by $\times$ symbols, while open circles represent the total relative uncertainty, $\Delta P_2/P_2 = \sqrt{\Delta P_{2,\mathrm{m}}^2 + \Delta P_{2,\mathrm{r}}^2}/P_2$.
Values of the total uncertainty are below $10\,\%$ for helium and $13\,\%$ for nitrogen, consistent with the uncertainties of the used commercial gauges.
Contributions from our sensor's random errors, $\Delta P_{2,\mathrm{r}}/P_2 \sim 1\,\%$, are negligible.
The larger deviation observed for nitrogen may be related to the limited temperature range over which the TS parameters $A$, $B$, and $C$ are defined; in Ref.~\cite{takaishi1963thermal}, these parameters are given for temperatures between $77$ and $195$~K, whereas the upper temperature in our setup is $T_1 = 295$~K.

\begin{figure}
\includegraphics[width=\linewidth]{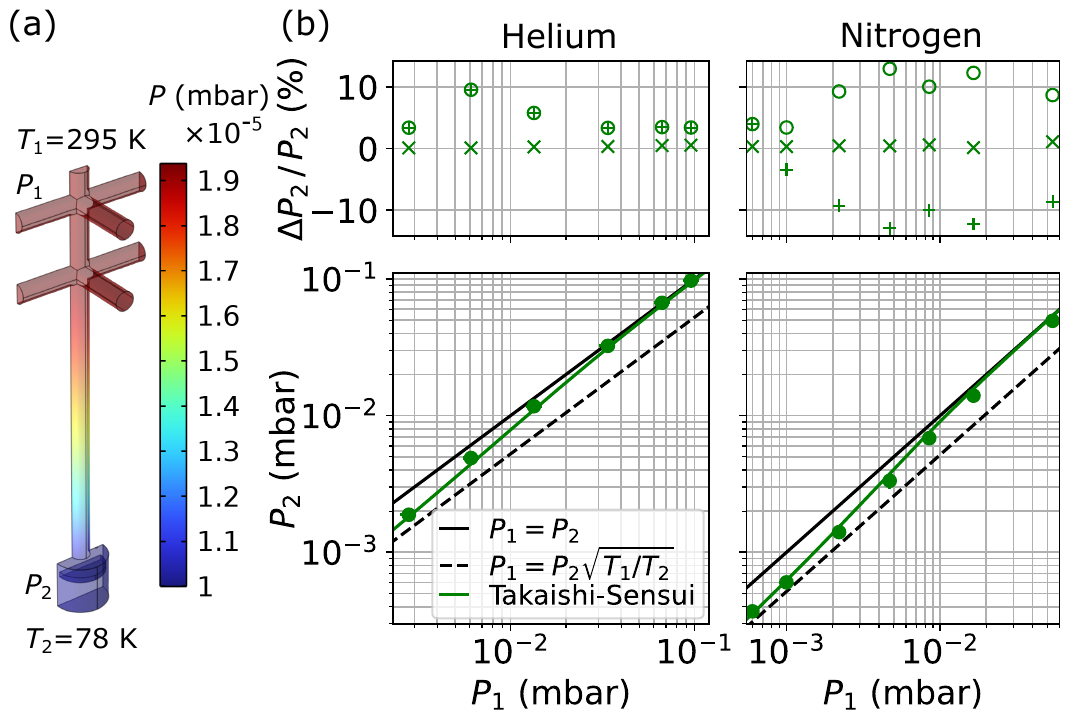}
\caption{Pressure measurements without active gas pumping or injection. (a) Simulated helium pressure distribution inside the vacuum chamber [see Fig.~\ref{fig:cryostat}(a)] at $T_1 = 295$~K and $T_2 = 78$~K in the absence of gas injection or pumping. 
Symmetry boundary conditions are applied to model one half of the chamber; a similar distribution is obtained for nitrogen (not shown).
The resulting pressure gradient originates from thermal transpiration and is consistent with the free-molecular-flow expectation $P_2/P_1 = \sqrt{T_1/T_2} = 1.94$. \\
(b, lower) Measured relations between $P_1$ and $P_2$ (green dots) for helium (left) and nitrogen (right) under the same conditions. 
The corresponding predictions of the Takaishi–Sensui (TS) model are shown as solid green lines, smoothly transitioning between the free-molecular limit ($P_2/P_1 = \sqrt{T_1/T_2}$, dashed line) and the viscous limit ($P_2 = P_1$, solid black line). \\
(b, upper) Quantities plotted relative to the measured $P_2$: the deviation between measured values and the TS model ($+$), the random measurement error ($\times$), and the root-sum-square of these two contributions ($\circ$); see text for details.
}
\label{fig:78K_pump_off}
\end{figure}

\begin{figure}
\includegraphics[width=\linewidth]{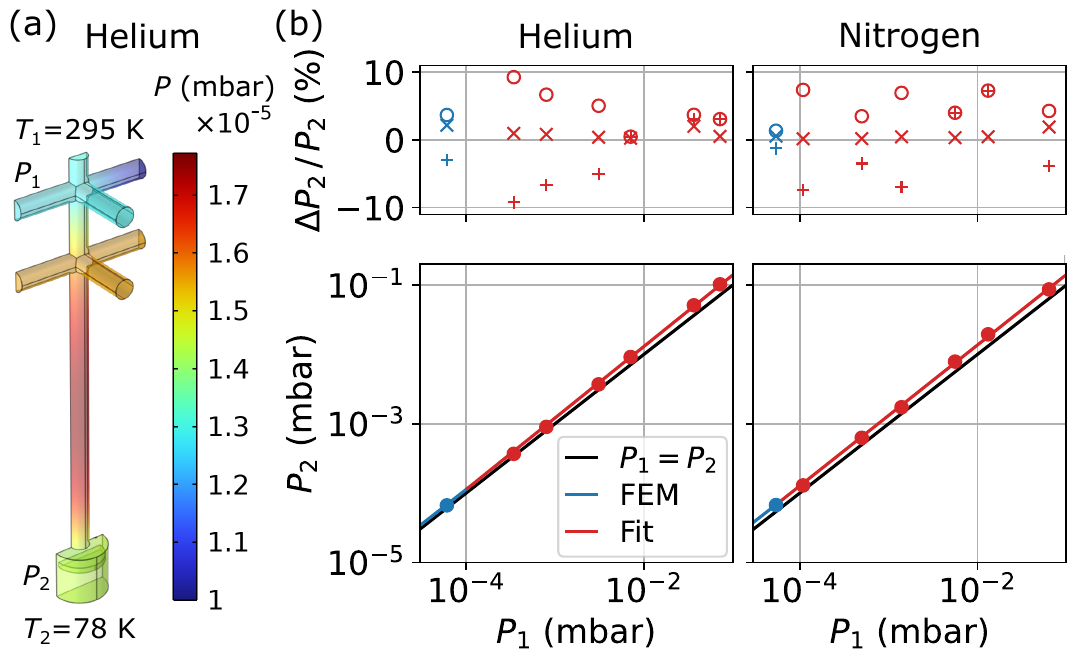}
\caption{Pressure measurements with active gas pumping and injection.
(a) Simulated helium pressure distribution inside the vacuum chamber [see Fig.~\ref{fig:cryostat}(a)] at $T_1 = 295$~K and $T_2 = 78$~K (half of the chamber simulated using symmetry boundary conditions). A local pressure maximum appears in the connecting pipe due to the combination of upward-directed gas flow from injection and opposing pressure increase in the upper volume from thermal transpiration. Similar distributions are obtained for nitrogen (not shown).\\
(b, lower) Measured and modeled relations between $P_1$ and $P_2$ for helium (left, $T_2 = 78$~K) and nitrogen (right, $T_2 = 85$~K) at $T_1 = 295$~K. Finite-element (FEM) results for the free-molecular flow regime and a two-parameter fit at higher pressures are shown for comparison. Data points are color-coded for molecular-flow (blue) and higher-pressure (red) regimes.\\
(b, upper) Deviation between measurements and models ($+$), random measurement error ($\times$), and root-sum-square of these contributions ($\circ$), all expressed relative to measured $P_2$ values; see text for details.}
\label{fig:78K_85K_pump_on}
\end{figure}

In a second step, to investigate our sensor at lower pressures, $P_2<6\times10^{-4}$~mbar, we pump on the warm end of the vacuum chamber, while injecting gas into the cold volume.
Figure~\ref{fig:78K_85K_pump_on}(a) shows the helium pressure distribution inside the vacuum chamber, simulated in COMSOL Multiphysics, at $T_1=295$~K and $T_2=78$~K. 
In contrast to the monotonic pressure gradient obtained without injecting and pumping gas (Fig.~\ref{fig:78K_pump_off}), a local pressure maximum appears inside the connecting pipe.
This is a consequence of two counteracting effects: the gas flow from lower to upper volume, due to injecting and pumping gas, respectively, and the temperature gradient which causes an increased pressure in the upper volume.
The simulated ratio of cold to warm volume pressures is $P_2/P_1=1.13$ for helium and $P_2/P_1=1.26$ for nitrogen, with a pressure distribution (not shown) similar to that of helium. 

The lower panels of Fig.~\ref{fig:78K_85K_pump_on}(b) show the measured and modeled relations between $P_1$ and $P_2$ for helium (left) and nitrogen (right), with the lower volume of the vacuum chamber cooled to 78~K and 85~K, respectively, while the upper volume is maintained at 295~K. 
In the free molecular flow regime, the measured relations between $P_1$ and $P_2$ for helium and nitrogen (blue dots) are compared to the COMSOL simulations (blue line). 
At higher pressures, measurements (red dots) are compared to a two-parameter fit (red line), $P_2 = a P_1^b$, with $a = 1.50$ and $b = 1.03$ for helium, and $a = 1.47$ and $b = 1.02$ for nitrogen; the fit connects continuously to the COMSOL model. 
In the upper panels of Fig.~\ref{fig:78K_85K_pump_on}(b), the symbols $+$, $\times$, and $\circ$ represent the measurement-to-model deviations, the random measurement errors, and the RSS of these two contributions, respectively, all given relative to $P_2$, as described in detail for Fig.~\ref{fig:78K_pump_off}(b).  
The RSS value is within $10\,\%$ for helium and $8\,\%$ for nitrogen, with the model-to-measurement deviation being the dominant contribution, while the random measurement error, which is up to $2\,\%$, contributes marginally.

\section{Conclusion}
We demonstrate a centimeter-sized gas pressure sensor that combines a nanomechanical membrane with a fiber-based laser interferometric readout, operating reliably down to 78~K. 
Quantitative pressure measurements in a confined 0.7~L cryogenic volume, spanning from $5\times10^{-5}$ to $10^{-1}$~mbar, show excellent agreement with the Takaishi–Sensui model for thermal transpiration and finite-element simulations in the free-molecular-flow limit. 
Since no commercial pressure sensors are available for direct cryogenic comparison, we compared our sensor to commercial gauges at room temperature using corresponding models. 
Relative uncertainties remain below $10\,\%$ for helium and $13\,\%$ for nitrogen, with minimal random errors, in agreement with the uncertainties of our commercial gauges.

The fiber-based interferometric readout achieves a displacement sensitivity of $S_x = 8\times10^{-14}$~m/$\sqrt{\mathrm{Hz}}$, currently limited by photodetector noise and non-optimal fiber-membrane alignment. 
This sensitivity is already sufficient to resolve thermal displacement noise of trampoline membrane resonators at atmospheric pressure, matching the performance achieved with free-space optical readout ~\cite{reinhardt2024self}.
Optical modeling suggests that a straightforward alignment optimization could further increase readout sensitivity towards shot-noise-limited operation at $10^{-14}$~m/$\sqrt{\mathrm{Hz}}$. 

These results demonstrate the successful transition of nanomechanical-membrane gas pressure sensing with lab-scale experiments to broadly-applicable portable devices, compatible with the previously demonstrated features, such as the 10-decade measurement range, gas-type-independent response, and self-calibrating operation \cite{reinhardt2024self,salimi2024squeeze,green2025accurate}.

Future work will focus on reducing alignment tolerances for optimized sensitivity, using patterned membranes with lower intrinsic loss ~\cite{reinhardt2016ultralow,tsaturyan2017ultracoherent,bereyhi2022hierarchical} to extend the measurement range; and incorporating squeeze-film coupling for improved sensitivity and the additional capability to measure the average mass of the gas particles \cite{EP4446714B1}.
Integrating temperature sensing, as well as extending the operational range to lower temperatures and higher pressures, will further broaden the sensor’s capabilities.

Overall, these developments pave the way for ultra-compact pressure sensors with unprecedented measurement range, sensitivity, and accuracy, extending their applicability from conventional vacuum technology to quantum technologies, precision metrology, and fundamental physics experiments.

\section{Acknowledgements}
We thank Friederike Januschek, Sven Karstensen, Jörn Schaffran, Antonio de Zubiaurre-Wagner, and the late Dieter Trines for helpful discussions. 
This work was supported and partly financed by the DESY Generator Program (D.H.), a PIER Seed Project PIF-2021-08, and by the DFG under Germany’s Excellence Strategy EXC 2121 “Quantum Universe”-390833306 (H.M.).

\appendix
\section{Plane-wave spectrum calculation}
\label{appsec:plane_wave_spectrum}

This appendix describes the reflected signal in a fiber interferometer using the angular spectrum (plane-wave) representation.
Following the approach of Ref.~\cite{li_theoretical_2006} for a Gaussian beam reflected by a dielectric slab, the formulation includes beam divergence and angular misalignment (see Fig.~\ref{fig:optics_sim}). 

To account for oblique incidence on the membrane, the plane-wave spectrum and direction of propagation are rotated. 
The wave vector components in medium $i$ are then given by
\begin{align}
    k^\prime_x(\theta) &= k_x \cos\theta - k_z \sin\theta, \\
    k^\prime_z(\theta) &= k_x \sin\theta + k_z \cos\theta, 
    \label{eqn:rot}
\end{align}
where the longitudinal component in medium $i$ is
\begin{align}
    k_{z,i} = \sqrt{n_i^2k_0^2 - k_x^2 - k_y^2}.
\end{align} 

The rotated angular spectrum corresponding to the Gaussian field at the fiber output is
\begin{equation}
    \Psi_0(k_x,k_y,\theta)
    = \sqrt{\frac{w_0^2}{2\pi}}
      \exp\!\biggl[-\frac{w_0^2}{4}\!\left(k_x^{\prime 2}(\theta) + k_y^2\right)\!\biggr],
    \label{eqn:spec_rot}
\end{equation}
with waist radius $w_0$ and vacuum wavenumber $k_0=2\pi/\lambda$.

\subsection{Reflection and transmission coefficients}
\label{appsec:reflection_coeffs}
For an interface between media $1$ and $2$, the Fresnel reflection and transmission coefficients for s- and p-polarized components are
\begin{align}
    r_{12}^s &= \frac{k^\prime_{1z} - k^\prime_{2z}}{k^\prime_{1z} + k^\prime_{2z}}, &
    t_{12}^s &= \frac{2k^\prime_{1z}}{k^\prime_{1z} + k^\prime_{2z}}, \\
    r_{12}^p &= \frac{n_2^2k^\prime_{1z} - n_1^2k^\prime_{2z}}{n_2^2k^\prime_{1z} + n_1^2k^\prime_{2z}}, &
    t_{12}^p &= \frac{2n_1n_2k^\prime_{1z}}{n_2^2k^\prime_{1z} + n_1^2k^\prime_{2z}},
    \label{eqn:Fresnel_oblique}
\end{align}
where $k^\prime_{z,1}$ is given by Eq.~\eqref{eqn:rot} and
\begin{equation}
    k^\prime_{z,2} = \sqrt{k_2^2 - k_x^{\prime 2} - k_y^2}.
\end{equation}
Here, the index $s$ ($p$) denotes polarization perpendicular (parallel) to the plane of incidence ($xz$, according to Fig.~\ref{fig:optics_sim}).

For a dielectric slab of thickness $d$ and refractive index $n_3$, surrounded by medium $n_2$, the reflection coefficients for s and p polarizations are
\begin{align}
    r_{d}^{s/p} = 
    \frac{r_{23}^{s/p} + r_{32}^{s/p} e^{2 i k_{3z} d}}
         {1 + r_{23}^{s/p} r_{32}^{s/p} e^{2 i k_{3z} d}},
\end{align}
where $r_{ij}^{s/p}$ are the single-interface Fresnel coefficients given by Eq.~\eqref{eqn:Fresnel_oblique} and $k_{3z}$ is the longitudinal component inside the slab.

\subsection{Weighting for oblique incidence}

At oblique incidence, the field in the laboratory frame (fixed transverse polarization) is a superposition of local s- and p-components.
The effective scalar reflection and transmission coefficients used in the plane-wave synthesis are obtained by
\begin{align}
    \rho_{d}^{s/p} &= r_{d}^{s/p} 
       - \bigl(r_{d}^{s/p} - r_{d}^{p/s}\bigr)\frac{k_y^2}{k_x^2 + k_y^2}, \\
    \tau_{d}^{s/p} &= t_{d}^{s/p} 
       - \bigl(t_{d}^{s/p} - t_{d}^{p/s}\bigr)\frac{k_y^2}{k_x^2 + k_y^2}.
\end{align}
Analogous weighting is applied at the fiber--air interface, giving the effective coefficients $\rho_{12}$ and $\tau_{12}$.
These expressions correspond to the ``$-$'' weighting convention of Sec.~\ref{appsec:reflection_coeffs}, which preserves correct limits as $k_y\!\to\!0$.

\subsection{Spectrum propagation and multiple reflections}

Fourier transforming Eq.~\eqref{eqn:spec_rot} yields a Gaussian beam propagating at an angle $\theta$ toward the dielectric slab, as shown in Fig.~4 of Ref.~\cite{li_theoretical_2006}. 
The waist of the incident beam is located a distance $L_0$ from the slab, along the negative $z$ direction.
A fraction of this light is reflected by the slab and returns toward the fiber, where part of it couples back into the guided mode and another part reflects again toward the slab. 
For the $j$-th roundtrip, the beam center on the slab is displaced by
\begin{equation}
    x_j = x_{j-1} + \Delta x_j, \qquad
    \Delta x_j = 
    L_{2j-1}\sin\theta_{j-1} + L_{2j}\sin\theta_j,
\end{equation}
with $x_0=0$, propagation angle $\theta_j = \theta(2j+1)$, and optical path
\begin{equation}
    L_q = L
    \prod_{p=0}^{q}\frac{\cos\!\left[(p-1)\theta\right]}
                         {\cos\!\left[(p+1)\theta\right]}.
\end{equation}

The displaced spectrum at its waist is
\begin{align}
    \Psi_j(k_x,k_y,\theta_j)
    = \Psi_0(k_x,k_y,\theta_j)\,
      \exp(i k_x x_j),
\end{align}
and the accumulated phase up to the slab is
\begin{align}
    S_j(k_x,k_y,\theta_j)
    &= S_{j-1}\,
       \exp\!\left[-i k_z^\prime(\theta_j)
       \bigl(L_{2j-1}+L_{2j}\bigr)\right], \\
    S_0(k_x,k_y,\theta)
    &= \exp\!\left[-i k_z^\prime(\theta)L_0\right].
\end{align}

\subsection{Interference spectrum and coupled field}

Truncating to the first two reflected beams ($j=0,1$) and omitting the explicit polarization superscripts, the field spectrum at the fiber facet is
\begin{equation}
    \Psi_\mathrm{inter}
    \approx
    \tau_{21}\rho_{d}\tau_{12}\,e^{-i k_z z_f(x)}
    \Bigl[
      S_0\Psi_0
      + \rho_{d}\rho_{21}S_1\Psi_1
    \Bigr],
\end{equation}
where $z_f(x)$ represents the local height of the tilted fiber--air interface.

The portion coupled back into the guided mode is found by projecting onto the fundamental fiber mode,
\begin{equation}
    \Psi_\mathrm{fib}
    = \iint \Psi_0^*(k_x,k_y)\,\Psi_\mathrm{inter}(k_x,k_y)\,dk_x\,dk_y,
\end{equation}
and the total reflected power, accounting for interference between the back-coupled and directly reflected fields at the fiber facet, is
\begin{equation}
    P_\mathrm{r}
    = P_\mathrm{in}
      \iint\!\left|
      \rho_{12}\Psi_0 + \Psi_\mathrm{fib}
      \right|^2 dk_x\,dk_y.
\end{equation}

\subsection{Plane-wave limit}

In the limit of vanishing divergence and tilt, $L,\theta\to0$, the reflected power approaches the standard plane-wave Fabry–Pérot expression,
\begin{equation}
\label{eq:plane_wave_limit}
    P_\mathrm{r}
    = P_\mathrm{in}
      \!\left[
      1 + 
      \frac{r_d^2 + r_{12}^2 - r_d^2 r_{12}^2 - 1}
           {1 + r_d^2 r_{12}^2 - 2r_d r_{12}\cos(2kL)}
      \right].
\end{equation}


\section{COMSOL simulations of pressure distributions}\label{app:comsol}
In the free molecular flow regime, the pressure distribution inside the vacuum chamber [Fig.~\ref{fig:cryostat}(a)] is simulated using COMSOL. 

\begin{figure}[h]
\includegraphics[width=0.85\linewidth]{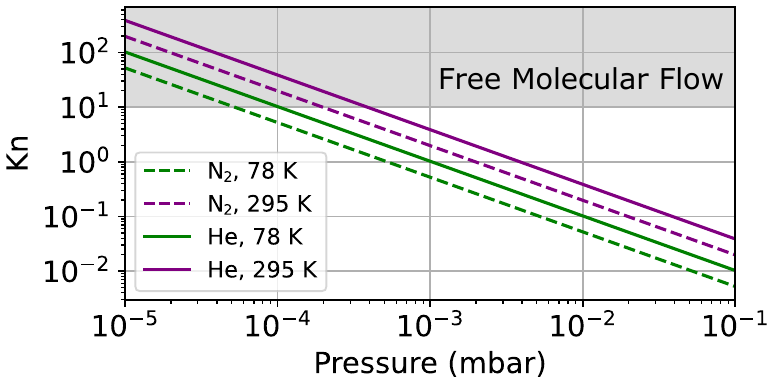}
\caption{Knudsen number $\text{Kn}$ as a function of pressure for N$_2$ (dashed) and He (solid) at 295~K (purple) and 78~K (green). 
Free-molecular flow occurs for $\text{Kn} > 10$.}
\label{fig:Knudsen}
\end{figure}

Figure~\ref{fig:Knudsen} shows the pressure dependency of the Knudsen number $\mathrm{Kn} = \ell/D$ ($\ell:$ mean free path of the gas particles, $D:$ pipe diameter of the vacuum chamber) for the used gases helium and nitrogen, at 295 K and 78 K, respectively. 
For $\mathrm{Kn}>10$ the system is in the free-molecular regime \cite{nakhosteen2016handbook}, which requires $P<10^{-4}\,...\,10^{-3}$, depending on gas type and temperature.

\begin{center}
\begin{table}[h]
\begin{tabular}{lcc}
\hline\hline
\textbf{Description} & \textbf{Parameter} & \textbf{Value} \\
\hline
Pumping speed for N$_2$ & $S_\mathrm{N_2}$ & 61~L/s \cite{Edwards} \\
Pumping speed for He & $S_\mathrm{He}$ & 57~L/s \cite{Edwards} \\
Angle valve conductance for N$_2$ & $C_\mathrm{v,N_2}$ & 45~L/s \cite{VAT} \\
Angle valve conductance for He & $C_\mathrm{v,He}$ & 121~L/s \cite{VAT} \\
Effective bellow length & $L_\mathrm{b,eff}$ & 0.7~m \\
Bellow diameter & $D$ & 0.035~m \\
\hline\hline
\end{tabular}
\caption{\label{tab:parameters}Parameters used to determine the effective pumping speed $S_\mathrm{eff}$ entering the COMSOL simulations of the pressure distribution inside the vacuum chamber.}
\end{table}
\end{center}

Based on the continuity of gas throughput $q$ between the lower and the upper parts of the vacuum chamber (Fig.~\ref{fig:cryostat}(a)), given by $q=C_\mathrm{p}\left(P_2-P_1\right)$, and through the pump, given by $q=S_\mathrm{eff}P_1$, one obtains $P_2 = \left(S_\mathrm{eff}/C_\mathrm{p}+1\right)P_1$ \cite{nakhosteen2016handbook}.
Here, $C_\mathrm{p}$ is the pipe conductance and $S_\mathrm{eff}=\left(S^{-1}+C^{-1}_\mathrm{v}+C^{-1}_\mathrm{b}\right)^{-1}$ is the effective pumping speed at the angle valve's inlet [see Fig.~\ref{fig:cryostat}(a)],with $S$ the nominal pumping speed and $C_\mathrm{v}$, $C_\mathrm{b}$ the conductances of the angle valve and bellow, respectively.   

The bellow conductance is estimated from its effective length \cite{krause2019improved}, which is approximately 17\,\% longer than its physical length, as
\begin{equation}
    C_\mathrm{b}=\frac{\pi D^3}{12L_\mathrm{b,eff}}\sqrt{\frac{8 k_\mathrm{B}T_\mathrm{gas}}{\pi m_\mathrm{gas}}}. 
    \label{eqn:flow_1}
\end{equation}
The parameters in Table~\ref{tab:parameters} result in $S_\mathrm{eff} = 13$~L/s for He and $6$~L/s for N$_2$.  
These values are used in the COMSOL simulations together with the detailed chamber geometry, stainless-steel material properties, and the gas-specific parameters.


%

\end{document}